\documentclass{aims}
\usepackage{amsmath,amssymb,comment}
  \usepackage{paralist}
  \usepackage{graphics} %% add this and next lines if pictures should be in esp format
  \usepackage{epsfig} %For pictures: screened artwork should be set up with an 85 or 100 line screen
\usepackage{graphicx}
 \usepackage{epstopdf}%This is to transfer .eps figure to .pdf figure; please compile your paper using PDFLeTex or PDFTeXify.
 \usepackage[colorlinks=true]{hyperref}
\hypersetup{urlcolor=blue, citecolor=red}
 \usepackage[mathlines,displaymath]{lineno}
  \def\currentyear{2022}
   \def\currentmonth{Feb.}
    \def\ppages{1--17}
     \def\DOI{10.3934/cpaa.2022029}

\newtheorem{theorem}{Theorem}[section]
\newtheorem{corollary}{Corollary}
\newtheorem*{main}{Main Theorem}
\newtheorem{lemma}[theorem]{Lemma}

\theoremstyle{definition}

\newcommand{\Beq}{\begin{equation}}
\newcommand{\Eeq}{\end{equation}}
\newcommand{\BS}{\begin{subequations}}
\newcommand{\ES}{\end{subequations}}
\newcommand{\Beqn}{\begin{equation*}}
\newcommand{\Eeqn}{\end{equation*}}
\newcommand{\Beqa}{\begin{eqnarray}}
\newcommand{\Eeqa}{\end{eqnarray}}
\newcommand{\Beqan}{\begin{eqnarray*}}
\newcommand{\Eeqan}{\end{eqnarray*}}

\title[Internal waves with Coriolis force]
      {Hamiltonian description of internal ocean waves with Coriolis force}

\author[Joseph Cullen and Rossen Ivanov]{}


 \email{joseph.d.cullen@icloud.com}
 \email{rossen.ivanov@tudublin.ie}

\thanks{R.I. is partially supported by the Bulgarian National Science Fund, grant K$\Pi$ -06H42/2
from 27.11.2020.
}

\begin{document}

\maketitle

\medskip
\centerline{\scshape Joseph D. Cullen and Rossen I. Ivanov}
\medskip
{\footnotesize
 % please put the address of the second  and third author
 \centerline{ School of Mathematical Sciences}
\centerline{ Technological University Dublin, City Campus}
\centerline{ Grangegorman Lower}
   \centerline{Dublin D07 ADY7, Ireland}
}

%\bigskip
%
%% The name of the associate editor will be entered by an editorial staff
%% "Communicated by the associate editor name" is not needed for special issue.
% \centerline{(Communicated by the associate editor name)}

%The abstract of your paper
\begin{abstract}
The interfacial internal waves are formed at the pycnocline or thermocline in the ocean and are influenced by the Coriolis force due to the Earth's rotation. A derivation of the model equations for the internal wave propagation taking into account the Coriolis effect is proposed. It is based on the Hamiltonian formulation of the internal wave dynamics in the irrotational case, appropriately extended to a {\it nearly} Hamiltonian formulation which incorporates the Coriolis forces.
Two propagation regimes are examined, the long-wave and the intermediate long-wave propagation with a small amplitude approximation for certain geophysical scales of the physical variables. The obtained models are of the type of the well-known Ostrovsky equation and describe the wave propagation over the two spatial horizontal dimensions of the ocean surface.
\end{abstract}

%The title of your section 1

\section{Introduction}

The dynamics of oceans and atmospheres in general has a strongly turbulent character, that is, the fluid motion is characterized by chaotic changes in the physical quantities, like the pressure and flow velocity with time. However, if the fluid motion is monitored
on large spatial and slow time scales, then patterns of highly ordered self-organized structures can emerge. These phenomena are
at the focus of geophysical fluid mechanics \cite{Ped} and among them are wave patterns of various nature.

The interfacial internal waves arise at the ``interface'' between a top layer of warm and less dense water over a lower layer of cold denser water. Both layers are to a big extent homogeneous of practicaly constant density and are separated by a sharp ``interface'' known as a thermocline (where the temperature gradient has a maximum) which is also very close to the pycnocline, (where the pressure gradient has a maximum) \cite{Ben1, Benj, HM, CaChoi, Osb}.  For the equatorial internal waves for example, the thermocline is usually at 100 - 200 m beneath the surface, \cite{FedorovBrown}. The internal ocean waves could have a significant impact on offshore engineering structures, such as oil platforms in the oceans as well as stationary tubes for oil and gas transportation stretching along the ocean shelf slope \cite{BV}.

The internal wave propagation is affected by various factors, such as currents and interactions with other waves, since the ocean dynamics near the surface is quite complex. However at great depths in the ocean (depths in excess of about 240 m) there is, essentially, an abyssal layer of still water. In our studies however we are not going to take into account these influences. It turns out that the highly idealised theoretical two-layer model describes quite accurately the reported observations \cite{Osb,HM}.

In this work we examine the Coriolis effect on the internal wave propagation following the  idea of ``nearly'' Hamiltonian approach, developed in series of previous papers like \cite{CIP07,CI2017,Compelli2,I19} and generalising the Hamiltonian approach of Zakharov \cite{Zak}. The aim of the paper is to illustrate the mathematical usefulness of the Hamiltonian approach in a systematic study of the internal wave propagation, rather than to present new equations on internal waves. Nevertheless the approach could be used in further studies, including detailed analysis with higher order approximations or effects, not included here.

The Coriolis effect is present both in the ocean and in the atmosphere. A mass of moving air or water subject only to the Coriolis force travels in a circular trajectory called an ``inertial circle''. The Coriolis effect has been intensively studied as well, see for example \cite{Ost,Leo,HM,Grim}.

For ocean waves of large magnitude, the viscosity does not play an essential role and can be neglected, so effectively the fluid dynamics is govern by Euler's equation.

\begin{figure}[ht]
\centering
\includegraphics[width=3.5in]{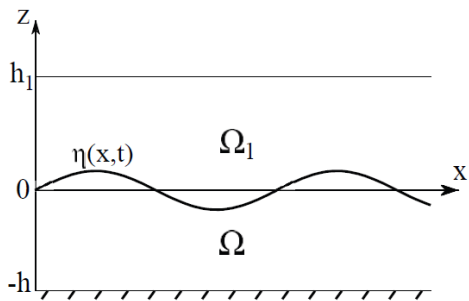}
    \caption{ System with an internal wave. The fluid domain $\Omega$ is the fluid of higher density. The pycnocline/thermocline
    is the interface that separates the two fluid domains $\Omega$ and $\Omega_1$. The function $\eta(x,t)$ describes the elevation of the internal wave.}
\label{figure1}
\end{figure}

\section{Internal waves with Coriolis force - the setup}
The Euler equation with included Coriolis force is
\begin{linenomath}
\begin{equation} \label{E}
{\bf V}_t+ ({\bf V} \cdot \nabla){\bf V})+ 2 \overrightarrow{ \omega} \times {\bf V}=-\frac{1}{\rho}\nabla p \end{equation} \end{linenomath}
where the velocity vector field ${\bf V}=(u,v,w)$ is presented through its components in a local coordinate system where the geophysical axis $x$ is oriented to the East, the $y$ axis is pointing to the North and the $z$ axis is vertical to the Earth surface. In addition we have incompressibility, i.e. $\text{div} {\bf V}=0.$  $p$ is the pressure in the fluid. The Earth's angular velocity at latitude $\theta$ in this system is \begin{linenomath} $$  \overrightarrow{ \omega} =\omega(0, \cos \theta, \sin \theta),$$ \end{linenomath} $\omega=7.3\times 10^{-5}$ rad/s. Introducing the parameters $f=2\omega \sin \theta$ and $r=2 \omega \cos \theta$ we have \begin{linenomath} $$ 2  \overrightarrow{ \omega} \times {\bf V}=(r  w  - f v, f u,-r u). $$ \end{linenomath}

Concerning the form \eqref{E}, we should specify that it arises as the so-called {\it traditional approximation} of the Euler equations in
spherical coordinates - see the discussion in \cite{CJ1}. It is also worth pointing out that at leading order the oblateness
of the Earth can be ignored even for atmospheric flows (and even more so for ocean flows) — see the discussion in the papers \cite{CJ2,CJ3}.

For Equatorial motion $\theta=0$ and $f=0$ so there are no forces acting in the $y$- direction. Moreover, the Coriolis forces are supporting the fluid to move along the Equator (in the $x$-direction), so that its motion remains two-dimensional.
Such situation with internal equatorial waves and currents is studied in \cite{CI2017}. We are going to consider now $\theta>0.$ In addition we assume that the fluid motion is irrotational (i.e. absence of currents and vorticity), apart from the global rotation caused by the Coriolis forces. In this approximation the velocity field is potential, i.e. ${\bf V}=\nabla \varphi(x,y,z,t)$ (hence $\Delta \varphi=0$) and the Coriolis effect will be presented as a perturbation to the potential motion.

The governing equations \eqref{E} acquire the form:
\begin{equation}
\begin{split}
& \left( \varphi_t + \frac{|\nabla \varphi|^2}{2}+ \frac{p}{\rho}+ gz \right)_x +r \varphi_z - f\varphi_y=0, \\
&\left( \varphi_t + \frac{|\nabla \varphi|^2}{2}+ \frac{p}{\rho}+ gz \right)_y + f\varphi_x=0, \\
&\left( \varphi_t + \frac{|\nabla \varphi|^2}{2}+ \frac{p}{\rho}+ gz \right)_z -r \varphi_x =0,
\end{split} \label{E2}
\end{equation}
where $g$ is the Earth acceleration. The internal waves are illustrated on Fig. \ref{figure1}. For fixed $y$ the system is bounded at the bottom by an impermeable flatbed and is considered as being bounded on the top by a flat horizontal surface, reflecting the assumption of absence of surface motion. The domains
\begin{equation} \nonumber
    \begin{split}
        &\Omega=\{(x, z)\in\mathbb{R}^2: -h< z < \eta(x,t)\} ,\\
        &\Omega_1=\{(x, z)\in\mathbb{R}^2: \eta(x,t)< z < h_1\}
    \end{split}
\end{equation}
corespond to the two fluid layers and the quantities associated with each domain are using the respective subscript notation. Also, subscript $c$ (implying \emph {common interface}) will be used to denote evaluation on the internal wave $z=\eta(x,t)$. Propagation of the internal wave is assumed to be in the positive $x$-direction, which is oriented eastward. The function $\eta(x,t)$ describes the elevation of the internal wave with the mean of $\eta$ assumed to be zero, \begin{equation} \label{C1}
    \int_{\mathbb{R}} \eta(x,t) dx=0.
\end{equation}

The system is considered incompressible with $\rho$ and $\rho_1$ being the respective constant densities of the lower and upper media and stability is given by the immiscibility condition $\rho>\rho_1.$ For long internal waves with wavelength $\lambda$ the parameter $\delta=h/\lambda \ll 1$ is a small non-dimensional parameter and $\varphi$ is a small quantity of order $\delta$, see for example \cite{CI2017}. The effect of the terms proportional to $r$ on the propagation in the $x$-direction could be estimated from the correction of the wave propagation speed in the $x$-direction due to the Coriolis force. The exact expression for this wave-propagation speed for equatorial waves when $\cos \theta =1$ is \cite{CI2017}
\begin{equation}
    c_0=-\alpha_1 (\rho- \rho_1) \omega \pm \sqrt{\alpha_1^2 (\rho- \rho_1)^2 \omega ^2+  \alpha_1 (\rho- \rho_1)g} , \quad \alpha_1=\frac{h h_1}{\rho_1 h + \rho h_1},
\end{equation}
The terms with $\omega$ (which are proportional to $r$) are much smaller, thus the approximate value of $c_0$
is $c_0=\pm \sqrt{\frac{h h_1(\rho-\rho_1)g}{(\rho_1h+\rho h_1)}}$ and the relative change due to the Coriolis force could be estimated to be of a magnitude  \begin{linenomath}$$\left| \frac{\Delta c_0}{c_0} \right|=\sqrt{\frac{h h_1(\rho-\rho_1)}{(\rho_1h+\rho h_1)g}} (2\omega \cos \theta )\sim 10^{-5} \,\, \text{to} \,\, 10^{-4}$$ \end{linenomath} for the feasible values of the parameters. Thus the dependence of the motion in the $(x,z)$-plane due to the terms containing $r$ could be neglected.

The motion in the $y$ direction is very slow in comparison to the wave propagation in the $x$-direction, therefore in leading order we have $p=p(x,z)$ and we can use the second equation in  \eqref{E2} in linear approximation to exclude the $y$ dependence,
 \begin{equation} \label{y-dep}
     \varphi_{ty}+f \varphi_x=0
 \end{equation}
  giving formally \begin{equation} \varphi_{y}=-f \partial_t^{-1} \varphi_x. \nonumber \end{equation}

  In what follows it will be assumed that $f$ is of order $\delta^{3/2}$ or $\delta^2 \ll 1.$ Noting that the $\partial_x$ operator with an eigenvalue $k=2\pi/\lambda$ is also of order $\delta$  (since $k=2\pi \delta/h$), for compatible time-scales $\partial_t \sim \delta$ thus the $y$-derivative $\varphi_y \ll \varphi_x.$ The first equation from \eqref{E2} gives the following generalisation of the Bernoulli equation:
\begin{equation} \varphi_t + \frac{|\nabla \varphi|^2}{2}+ \frac{p}{\rho}+ gz + f^2\partial_t^{-1} \varphi=0. \nonumber \end{equation}
The nonlinear contribution $|\nabla \varphi|^2=\varphi_x^2+\varphi_y^2+\varphi_z^2\approx \varphi_x^2+\varphi_z^2. $
Therefore, the quantities in the leading order do not depend on the $y$ variable and in the following sections we will suppress this dependence and will consider essentially two-dimensional wave motion (the $y$-dependence will be discussed again and reintroduced in Section \ref{y-depe}). We point out that the two-dimensional motion is common at the Equator ($\theta=0$), but not in most other parts of the ocean where meandering is quite frequent. Nevertheless it is also physically realistic when $\theta>0:$ perhaps the best example for such two-dimensional wave propagation is the Antarctic Circumpolar Current - see the discussion in \cite{CJ4,HaMa}. In this case the Bernoulli equation
\begin{equation} \varphi_t + \frac{ \varphi_x^2+ \varphi_z^2}{2}+ \frac{p(x,z)}{\rho}+ g z + f^2\partial_t^{-1} \varphi=0. \label{IW} \end{equation}
effectively becomes a (2+1)-dimensional equation  for the $x,z$ dependent variables (considering $y$ fixed) and it will provide one of the governing equations for the arising models.

We make the assumption that the functions $\eta(x, t),$ ${\varphi}(x, z, t)$ and ${\varphi}_1(x, z, t)$ belong to the Schwartz class $\mathcal{S}(\mathbb{R})$ with respect to the $x$ variable (for any $z$ and $t$). This reflects the localised nature of the wave disturbances, that is disturbances in the form of solitary waves. The assumption of course implies that for large absolute values of $x$ the internal wave attenuates \begin{equation}
\lim_{|x|\rightarrow \infty}\eta(x,t)=0, \quad \lim_{|x|\rightarrow \infty}{ {\varphi}}(x,z,t)=0 \quad\mbox{and} \quad \lim_{|x|\rightarrow \infty}{ {\varphi}_1}(x,z,t)=0.
\end{equation} The action of the operators $\partial_x ^{-1},$ is not uniquely defined, however for the special subclass of functions satisfying the condition \eqref{C1} the action is unique,
\begin{equation}
  \partial_x ^{-1}\eta(x, t)=  \int_{\pm\infty} ^x \eta(x',t)dx',
\end{equation}
and moreover $\partial_x ^{-1}\eta(x, t)$ belongs to $\mathcal{S}(\mathbb{R})$ with respect to the $x$ variable as well.
Therefore, the differentiation operator is invertible on the functional subclass of $\mathcal{S}(\mathbb{R})$ satisfying \eqref{C1}.
For a slightly wider functional classes, like $\eta(x,\cdot) \in \mathcal{S}(\mathbb{R})$, we extend the solution set by allowing {\it weak} solutions, such that formal expressions like $P[\eta, \eta_x, \ldots]=\partial_t^{-1}Q$ have the only meaning that $P_t=Q$ and are used only for convenience, not for the explicit determination of $P.$ An analogous remark holds if $P=\partial_x^{-1}Q$ and $Q$ does not satisfy \eqref{C1}.

\section{(Nearly) Hamiltonian representation of the internal wave dynamics}

Recall that we use subindex $1$ for the quantities, associated to the upper layer. Also, subscript $c$ (implying \emph {common interface}) will be used to denote evaluation on the internal wave $z=\eta(x,t)$. In addition, $\eta(x,t)$ satisfies the boundary kinematic condition on the interface
\begin{equation}\label{bcc}
  \eta_t=(w -u \eta_x)_c=(w_{1} -u_{1} \eta_x)_c
\end{equation}
or
\begin{equation}\label{bcc1}
  \eta_t=(\varphi_z -\varphi_x \eta_x)_c=(\varphi_{1,z} -\varphi_{1,x} \eta_x)_c
\end{equation} representing the fact that the particles from the surface $z=\eta(x,t)$ (which is actually the interface, also the thermocline and the pycnocline) do not move in the direction, perpendicular to the surface \cite{Johnson_Book}.

Propagation of the internal wave is assumed to be in the positive $x$-direction which is considered to be ``eastward''.  The main result is formulated in the following theorem:

\begin{main} The time evolution of the internal wave motion in the $x$-direction is given by the following system in quasi-Hamiltonian form
\begin{equation} \label{H1}
\xi_t=-\frac{\delta H_0}{\delta \eta}- f^2 \left(\partial_t^{-1} (\rho \varphi -\rho_1 \varphi_1)  \right)_c
\end{equation}
\begin{equation} \label{H2}
\eta_t=\frac{\delta H_0}{\delta \xi}.
\end{equation} where \begin{equation}\xi(x,t):= (\rho \varphi-\rho_1 \varphi_1)_c. \nonumber \end{equation} and $H_0$ is the Hamiltonian that corresponds to the irrotational motion ($f=0$).

\end{main}

Proof: At $z=\eta(x,t)$ we have $p(x,\eta,t)=p_1(x,\eta,t)$ and therefore we have the Bernoulli condition
\begin{equation}
\rho \! \left( \! ({{\varphi}_{t}})_c+\frac{1}{2}|\nabla \varphi|_c^2 +g\eta + f^2 (\partial_t^{-1} \varphi)_c \! \right)=\rho_1 \! \left(\!( {{\varphi}_{1,t}})_c+\frac{1}{2}|\nabla \psi_1|_c^2 +g\eta + f^2 (\partial_t^{-1} \varphi_1)_c \! \right)
\end{equation}
or
\begin{equation}
\left(\rho \varphi_t-\rho_1 \varphi_{1,t}\right)_c+\frac{\rho}{2}|\nabla \varphi|_c^2 - \frac{\rho_1}{2}|\nabla \varphi_1|_c^2   +g(\rho-\rho_1)\eta + f^2 \left(\partial_t^{-1} (\rho \varphi -\rho_1 \varphi_1)  \right)_c=0. \label{Ber}
\end{equation}

The functional $H_0$, which describes the total energy of the system, can be written as the sum of the kinetic, $\mathcal{K}$, and potential energy, $(\Pi)$ contributions. The potential part, must be
\begin{equation}\mathcal{V}(\eta)=\rho g\int_{\mathbb{R}} \int_{h}^{\eta}  z \, dz dx +\rho_1 g\int_{\mathbb{R}} \int_{\eta}^{h_1}  z \, dz dx.\end{equation} However, the potential energy is always measured from some reference value, e.g. $\mathcal{V}(\eta=0)$ which is the potential energy of the undisturbed interface, without wave motion. Therefore, the relevant part of the potential energy, contributing to the wave motion is  \begin{equation} \Pi(\eta)= \mathcal{V}(\eta)-\mathcal{V}(0)=\rho g\int_{\mathbb{R}} \int_{0}^{\eta}  z \, dz dx +\rho_1 g\int_{\mathbb{R}} \int_{\eta}^{0}  z \, dz dx=\frac{1}{2}(\rho-\rho_1)g\int_{\mathbb{R}} \eta^2 dx
.\end{equation}

The kinetic energy of the wave motion is
\begin{equation}
\mathcal{K}=\frac{1}{2}\rho\int_{\mathbb{R}} \int_{-h}^{\eta}  (u^2+w^2)dz dx+\frac{1}{2}\rho_1\int_{\mathbb{R}} \int_{\eta}^{h_1}  (u_1^2+w_1^2)dz dx
\end{equation} and in terms of the velocity potentials

\begin{equation}
\begin{split}
H_0=\mathcal{K}+ \Pi &=\frac{1}{2}\rho\int_{\mathbb{R}} \int_{-h}^{\eta}  |\nabla \varphi |^2dz dx+\frac{1}{2}\rho_1\int_{\mathbb{R}} \int_{\eta}^{h_1}
  |\nabla \varphi_{1}|^2 dz dx \\
  &+ \frac{1}{2}(\rho-\rho_1)g\int_{\mathbb{R}} \eta^2 dx.
\end{split}
\end{equation}
The next step is the computation of $\delta H_0:$
\begin{equation}\label{H0-var} \begin{split}
  \delta H_0&=\rho\int_{\mathbb{R}} \int_{-h}^{\eta}  \nabla \varphi \cdot \nabla (\delta \varphi) \, dz dx+\rho_1\int_{\mathbb{R}} \int_{\eta}^{h_1}
  \nabla \varphi_{1} \cdot \nabla (\delta \varphi_1) \, dz dx \\
  &+ \frac{1}{2}\rho\int_{\mathbb{R}}  |\nabla \varphi |^2_c \delta \eta \, dx - \frac{1}{2}\rho_1\int_{\mathbb{R}} |\nabla \varphi_{1}|^2 _c \delta \eta \, dx + (\rho-\rho_1)g\int_{\mathbb{R}} \eta \delta \eta \, dx.
  \end{split}
\end{equation}
Taking into account $\Delta \varphi =0$ and $\Delta \varphi_1 =0$ we have
\begin{equation}\label{H0-var1}
\begin{split}
  \delta H_0&=\rho\int_{\mathbb{R}} \int_{-h}^{\eta}  \mathrm{div }[(\nabla \varphi )\delta \varphi] \, dz dx+\rho_1\int_{\mathbb{R}} \int_{\eta}^{h_1}
  \mathrm{div } [(\nabla \varphi_{1})\delta \varphi_1] \, dz dx \\
  &+ \int_{\mathbb{R}} \left(\frac{\rho}{2}  |\nabla \varphi |^2_c -\frac{\rho_1}{2} |\nabla \varphi_{1}|^2 _c  +(\rho-\rho_1)g\eta  \right)\delta \eta \, dx  .
  \end{split}
\end{equation}
We apply the divergence Theorem for the domains $\Omega$ and $\Omega_1$, noticing that the outward normal to the domain $\Omega$ is $\mathbf{N}=(-\eta_x, 1),$ the outward normal to $\Omega_1$ is $-\mathbf{N},$ $|\mathbf{N}|=\sqrt{1+\eta_x^2},$ the line element is $dl = \sqrt{1+\eta_x^2}dx$ and there is no contribution from the flat bed and top surface:
\begin{equation}\label{H0-var2}
\begin{split}
  \delta H_0&=\rho\int_{\mathbb{R}} ( \varphi_z -\eta_x \varphi_x )_c (\delta \varphi)_c \, dx - \rho_1\int_{\mathbb{R}}
  (\varphi_{1,z}-\eta_x \varphi_{1,x})_c (\delta \varphi_1)_c \, dx \\
  &+ \int_{\mathbb{R}} \left(\frac{\rho}{2}  |\nabla \varphi |^2_c -\frac{\rho_1}{2} |\nabla \varphi_{1}|^2 _c  +(\rho-\rho_1)g\eta  \right)\delta \eta \, dx  .
  \end{split}
\end{equation}
Using \eqref{bcc} we obtain
\begin{equation}\label{H0-var3}
\begin{split}
  \delta H_0&=\int_{\mathbb{R}} \eta_t [\rho(\delta \varphi)_c -\rho_1(\delta \varphi_1)_c ]\, dx \\
  &+ \int_{\mathbb{R}} \left(\frac{\rho}{2}  |\nabla \varphi |^2_c -\frac{\rho_1}{2} |\nabla \varphi_{1}|^2 _c  +(\rho-\rho_1)g\eta  \right)\delta \eta \, dx  .
  \end{split}
\end{equation}
In addition, we have the relations $(\delta \varphi)_c =\delta \varphi_c -(\varphi_z)_c \delta \eta$ and \\
$(\delta \varphi_1)_c =\delta (\varphi_{1})_c -(\varphi_{1,z})_c \delta \eta,$ leading to
\begin{equation}\label{H0-var4}
\begin{split}
  \delta H_0&=\int_{\mathbb{R}} \eta_t \delta \xi \, dx \\
  &+ \int_{\mathbb{R}} \left( \frac{\rho}{2}  |\nabla \varphi |^2_c -\frac{\rho_1}{2} |\nabla \varphi_{1}|^2 _c  +(\rho-\rho_1)g\eta  -(\rho (\varphi_z)_c-\rho_1(\varphi_{1,z})_c)\eta_t \right)\delta \eta \, dx  .
  \end{split}
\end{equation}
From \eqref{H0-var4} we have immediately equation \eqref{H2} and also
\begin{equation}\label{deltaH_0}
 \frac{\delta H_0}{\delta \eta} = \frac{\rho}{2}  |\nabla \varphi |^2_c -\frac{\rho_1}{2} |\nabla \varphi_{1}|^2 _c  +(\rho-\rho_1)g\eta  -(\rho \varphi_z-\rho_1\varphi_{1,z})_c\eta_t.
\end{equation}
 Since  $( \varphi_c)_t =( \varphi_t)_c +(\varphi_z)_c  \eta_t$ and $(\varphi_{1,c})_t = (\varphi_{1,t})_c +(\varphi_{1,z})_c  \eta_t,$ then
 \begin{equation}\label{xit}
   \xi_t=(\rho \varphi_t-\rho_1 \varphi_{1,t})_c+(\rho \varphi_z-\rho_1\varphi_{1,z})_c\eta_t.
 \end{equation}
 From \eqref{deltaH_0} and \eqref{xit} it follows that
\begin{equation}\label{deltaH_0a}
 \frac{\delta H_0}{\delta \eta} = \frac{\rho}{2}  |\nabla \varphi |^2_c -\frac{\rho_1}{2} |\nabla \varphi_{1}|^2 _c  +(\rho-\rho_1)g\eta  +(\rho \varphi_t-\rho_1\varphi_{1,t})_c-\xi_t.
\end{equation}

Now \eqref{H1} follows from \eqref{deltaH_0a} and \eqref{Ber}.
Hence \eqref{H1} -- \eqref{H2} represent the nearly Hamiltonian formulation of the internal wave dynamics. \qed

The Hamiltonian $H_0[\xi,\eta]$ can be expressed as a functional of the variables $\xi, \eta$ with the help of the so-called Dirichlet-Neumann operators \cite{Craig,Craig2}. These operators do not have an explicit form, however, they admit series expansions over the  a small parameter of the system (such as, for example, small amplitude $|\eta_{\mathrm{max}}|/h \ll 1$).
In the following sections the Main Theorem will be applied in the derivation of several important approximate model equations under various assumptions for the smallness of the physical quantities.

\section{Long wave and small amplitude approximation}

In this approximation $\delta\ll 1,$  $\eta \sim \delta^2,$ $D=-i\partial_x \sim \delta,$ $\xi \sim \delta,$ and $f \sim \delta^2.$ The expansion with respect to $\delta$ of the Hamiltonian is (see \cite{CI2017})
\begin{equation} \label{H}
H_0[\xi,\eta]=\frac{\delta^4}{2} \int_{\mathbb{R}} \xi D\left( \alpha_1 + \delta^2 (\alpha_3 \eta - \alpha_2 D^2) \right)D \xi \,dx   + \delta^4 g(\rho-\rho_1) \int_{\mathbb{R}} \frac{\eta^2}{2} dx
 \end{equation}
 where
\begin{equation} \begin{split} \label{alphas}
 \alpha_1&=\frac{h h_1}{\rho_1 h+\rho h_1}, \qquad \alpha_2= \frac{h^2h_1^2(\rho h+\rho_1 h_1)}{3(\rho_1 h+\rho h_1)^2}, \qquad \alpha_3= \frac{\rho h_1^2-\rho_1 h^2}{(\rho_1 h+\rho h_1)^2}
\end{split}\end{equation} are constant parameters.

 The Hamiltonian equations that follow from \eqref{H1}-\eqref{H2} expressed in terms of $\eta$ and $\tilde{u}=\xi_x$ are
\begin{equation}\label{BA0}
\begin{split}
&\eta_t + \alpha_1 \tilde{u}_x + \delta^2 \alpha_2 \tilde{u}_{xxx} + \delta^2 \alpha_3 (\eta \tilde{u})_x =0, \\
&\tilde{u}_t+g(\rho-\rho_1)\eta_x + \delta^2 \alpha_3 \tilde{u}\tilde{u}_x + \delta^2 f^2\left[(\partial_t^{-1}(\rho \varphi-\rho_1 \varphi_1))_{z=\eta(x,t)}\right]_x =0.
\end{split}
\end{equation}
Let us analyse the following term:
\begin{equation}\partial_t^{-1}(\rho \varphi-\rho_1 \varphi_1))_{z=\eta(x,t)}= \int^t [\rho \varphi(x,\eta(x,t),t')-\rho_1 \varphi_1(x,\eta(x,t),t') ]dt'.
\nonumber \end{equation} We evaluate
\begin{linenomath}
\begin{multline}
\partial_t[\partial_t^{-1}(\rho \varphi-\rho_1 \varphi_1))_{z=\eta(x,t)}]=\\
[\rho \varphi(x,\eta(x,t),t)-\rho_1 \varphi_1(x,\eta(x,t),t)\! ]\! + \!\eta_t \!\int^t \![\rho \varphi_z(x,\eta(x,t),t')-\rho_1 \varphi_{1,z}(x,\eta(x,t),t') \!]dt'\\=\xi(x,t)+\text{smaller order terms}
\end{multline} \end{linenomath}
\noindent since $\xi \sim \delta$ and $\eta_t\sim \delta^3$  etc. Therefore
\begin{equation}\left[(\partial_t^{-1}(\rho \varphi-\rho_1 \varphi_1))_{z=\eta(x,t)}\right]_x=\partial_t^{-1}\xi_x+\ldots
=\partial_t^{-1}\tilde{u}+\ldots, \nonumber \end{equation}
which leads to the system of coupled equations
\begin{equation}\label{BA1}
\begin{split}
&\eta_t + \alpha_1 \tilde{u}_x + \delta^2 \alpha_2 \tilde{u}_{xxx} + \delta^2 \alpha_3 (\eta \tilde{u})_x =0, \\
&\tilde{u}_t+g(\rho-\rho_1)\eta_x + \delta^2 \alpha_3 \tilde{u}\tilde{u}_x + \delta^2 f^2(\partial_t^{-1} \tilde{u})  =0.
\end{split}
\end{equation}
In leading order \begin{equation}\label{BA3}
\eta_t + \alpha_1 \tilde{u}_x=0, \qquad  \tilde{u}_t+g(\rho-\rho_1)\eta_x=0
 \end{equation} or
\begin{equation}\label{BA4}
\eta_{tt}=- \alpha_1 \tilde{u}_{xt}= g\alpha_1(\rho-\rho_1)\eta_{xx}, \qquad \eta_{tt}-g\alpha_1(\rho-\rho_1)\eta_{xx}=0, \end{equation}
which is the wave equation for $\eta$ giving the wave speed
\begin{equation} \label{c_0}
c_0=\pm \sqrt{ \alpha_1(\rho-\rho_1)g}.
\end{equation}
For an observer, moving with the flow, i.e. there are left- ($-$ sign) and right-running ($+$ sign) waves. Moreover, in the leading approximation, for linear waves, the functions depend on the characteristic variable $x-c_0t$, therefore  $\tilde{u}=\frac{c_0}{\alpha_1} \eta.$ In the next order approximations with respect to $\delta$ obviously
\begin{equation} \tilde{u}=\frac{c_0}{\alpha_1} \eta + \delta^2 (\ldots) \label{JT}\end{equation} however we will not need this explicitly, see for example \cite{J02}. Differentiating the first equation in \eqref{BA1} with respect to $t$
\begin{equation}\eta_{tt} + \alpha_1 \tilde{u}_{tx} + \delta^2 \alpha_2 \tilde{u}_{txxx} + \delta^2 \alpha_3 (\eta_t \tilde{u}+ \eta \tilde{u}_t)_x=0  \nonumber \end{equation}
and substituting in it $\tilde{u}_t$ from the second equation in \eqref{BA1},
$\tilde{u}$ from \eqref{JT}, and \begin{equation}\eta_t=-\alpha_1 \tilde{u}_x + \delta^2(\ldots) \nonumber \end{equation} where necessary, neglecting $\delta^4$ terms, we obtain the following generalised Boussinesq equation for $\eta$:
\begin{equation}\label{Bo}
\eta_{tt}-c_0^2 \eta_{xx}-\delta^2\frac{3\alpha_3 c_0^2}{2\alpha_1}(\eta^2)_{xx}-\delta^2\frac{\alpha_2 c_0^2}{\alpha_1}\eta_{xxxx}+\delta^2 f^2 \eta=0
\end{equation}
The dispersion law of this equation is \begin{equation} \label{dl}
\tilde{\omega}^2(k)=c_0^2k^2 -\delta^2\frac{\alpha_2 c_0^2}{\alpha_1} k^4 +\delta^2 f^2 , \qquad \tilde{\omega}(k)\approx c_0k -\delta^2\frac{\alpha_2 c_0}{2\alpha_1} k^3 +\delta^2 \frac{f^2}{2kc_0}.\end{equation}
 Furthermore, a generalised Korteweg - de Vries (or KdV,  \cite{KdV}) type equation of the form
\begin{equation}\eta_t + c_0 \eta_x + \delta^2 a_1 \eta_{xxx}+ \delta^2 a_2 (\eta^2)_x + \delta^2 a_3 f^2 \partial_x^{-1 }\eta=0 \label{KdV1} \end{equation} for some constants $a_1,a_2,a_3$ (yet unknown) could be obtained from  \eqref{Bo}. Indeed, differentiating the above equation with respect to $t$ we have \begin{equation}\eta_{tt} + c_0 \eta_{xt} + \delta^2 a_1 \eta_{txxx}+ \delta^2 a_2 (\eta^2)_{xt} + \delta^2 a_3 f^2 \partial_x^{-1 }\eta_t =0 \nonumber \end{equation} in which we substitute $\eta_t$ from \eqref{KdV1} to obtain (neglecting $\delta^4$ terms)
\begin{equation}\eta_{tt}-c_0^2\eta_{xx}-\delta^2 2a_1 c_0 \eta_{xxxx} -\delta^2 2a_2 c_0(\eta^2)_{xx}-\delta^2 2 a_3 c_0 f^2 \eta =0. \nonumber \end{equation} The comparison with \eqref{Bo} gives
\begin{equation} a_1=\frac{\alpha_2 c_0}{2 \alpha_1}, \qquad a_2=\frac{3\alpha_3 c_0}{4\alpha_1},    \qquad a_3= - \frac{1}{2c_0}. \nonumber \end{equation}
 Then finally the KdV-type equation acquires the form
  \begin{equation}\eta_t + c_0 \eta_x + \delta^2 \frac{\alpha_2 c_0}{2 \alpha_1} \eta_{xxx}+ \delta^2 \frac{3c_0\alpha_3}{4\alpha_1} (\eta^2)_x = \delta^2 \frac{ f^2 }{2c_0}\partial_x^{-1 }\eta \label{OstrA} \end{equation}

The leading order term $\eta_t + c_0 \eta _x $ could be transformed to just a $T-$derivative via an appropriate Galilean transformation
$X\rightarrow (x-c_0t),$  $T\rightarrow \delta^2 t$  which introduces the characteristic variable $X$ and the slow time $T$ such that  $\eta_t =\delta^2 \eta_T-c_0 \eta_X,$ $\eta_x=\eta_X$ and $\eta_t + c_0 \eta_x=\delta^2 \eta_T$ and therefore
\begin{equation}\eta_T+  \frac{\alpha_2 c_0}{2 \alpha_1} \eta_{XXX}+ \frac{3 c_0 \alpha_3}{4\alpha_1} (\eta^2)_X =  \frac{ f^2 }{2c_0}\partial_X^{-1 }\eta \label{OstrB}. \end{equation}
In the chosen scaling all quantities are of the same order and the scale variable $\delta$ has disappeared.
The equation \eqref{OstrB} is also known as Ostrovsky's equation \cite{Ost}. Note that the dispersion law of the Ostrovsky equation \eqref{OstrA} is like in \eqref{dl}:
\begin{equation}\tilde{\omega}(k)= c_0k -\delta^2\frac{\alpha_2 c_0}{2\alpha_1} k^3 +\delta^2 \frac{f^2}{2kc_0}. \nonumber \end{equation}

The condition \eqref{C1} could be relaxed to $\int_{\mathbb{R}} \eta(x,t) dx$ = constant, which is typical for the soliton-like solutions. Such a condition does not change the average value of $\eta$ over the whole real line $\mathbb{R},$ since the interval in this case has an infinite length, and the average value is still zero. The inverse operators of $\partial_x$ and $\partial_X$ however then are not unique and we can use \eqref{OstrB} only in the form
\begin{equation}\left [\eta_T+  \frac{\alpha_2 c_0}{2 \alpha_1} \eta_{XXX}+ \frac{3c_0\alpha_3}{4\alpha_1} (\eta^2)_X \right ]_X=  \frac{ f^2 }{2c_0}\eta \label{OstrC}. \end{equation}

A derivation directly from Euler's equations is presented in Leonov's paper \cite{Leo}.

The Ostrovsky equation itself is Hamiltonian and possesses three conservation laws, however it is not bi-Hamiltonian and it is not integrable by the Inverse Scattering Method \cite{CIL07}. Solutions from perturbations of the KdV solitons \cite{ZMNP,Johnson_Book} can be derived in principle, although this is technically difficult, see for example \cite{GI} and the references therein.  Some special solutions like travelling waves for the Ostrovsky equation have been well known (see for example \cite{GOSS}) and can be used in the analysis of the Earth's rotation in processes like energy transfer \cite{GOSS,Henry} by ocean waves. Various other aspects of the equation have been studied extensively in numerous works, see for example \cite{GOSS,VL04} and the references therein.

\section{Special case of small or vanishing $\alpha_3$ }

There is one special configuration when the coefficient $\alpha_3$ from \eqref{alphas} is small (or vanishing), for example $ \mathcal{O}(\alpha_3) \le \delta,$  or $\rho h_1^2 \approx \rho_1 h^2.$ Then the nonlinearities like $\eta^2 \eta_x$ could contribute significantly in ballancing the dispersive terms. We can show that there is a  scaling which corresponds to this situation and it is $\eta \sim \delta,$  $f\sim \delta^{2}$ and, of course, $\tilde{u} \sim \delta.$ Because the scaling differs from the one in the previous section we present this case  separately.
The Hamiltonian acquires some extra terms. Looking at the Hamiltonian (5.9) from the article \cite{CI2017} and expanding up to terms of order $\delta^4$ we obtain
\begin{linenomath}
\begin{multline} \label{Ha}
H_0[\eta,\tilde{u}]=\frac{1}{2} \int_{\mathbb{R}}  \left(\delta^2 \alpha_1 + \delta^3\alpha_3 \eta \right)\tilde{u}^2 dx   +  \delta^2\int_{\mathbb{R}} \alpha_4\frac{\eta^2}{2} dx -\delta ^4 \frac{1}{2} \int_{\mathbb{R}} \beta_1 \eta^2 \tilde{u} ^2 dx
\\
+\delta^4 \frac{1}{2} \int_{\mathbb{R}}  \alpha_2 \tilde{u}\tilde{u}_{xx} dx,   \end{multline} \end{linenomath}
where $\alpha_4= g(\rho-\rho_1)$ and \begin{equation}\beta_1=\frac{\rho \rho_1(h+h_1)^2}{(\rho_1 h + \rho h_1)^3}. \nonumber \end{equation} The term with $\alpha_3$ is smaller as we know, it is of order $\delta^4$ or even smaller, but because $\alpha_3$ is a numerical coefficient and for some other choice of the parameters it could be significant we do not scale the coefficient $\alpha_3$ explicitly.
Taking into account the scaling, the ''nearly'' Hamiltonian formulation could be written in the form
\begin{equation} \label{H_Coriolis}
    \begin{split}
    \eta_t & =- [\delta^{-2}] \left(\frac{\delta H_0}{\delta \tilde{u}} \right)_x\\
       \tilde{u}_t & =- [\delta^{-2}] \left(\frac{\delta H_0}{\delta \eta} \right)_x- \delta^2 f^2 \partial_t^{-1} \tilde{u}
    \end{split}
\end{equation}
where $[\delta^{-2}] $ is a scale factor and not a variation. Explicitly we have the system of equations
\begin{linenomath}
\begin{eqnarray}
        \eta_t + \alpha_1 \tilde{u}_x+ \left(\delta \alpha_3 \eta \tilde{u} -\delta^2\beta_1 \eta^2 \tilde{u}+ \delta^2\alpha_2 \tilde{u}_{xx} \right)_x &=&0  \label{eq1}\\
       \tilde{u}_t +\alpha_4 \eta_x + \left(\delta\frac{\alpha_3 \tilde{u}^2}{2} - \delta^2\beta_1 \eta \tilde{u}^2\right)_x + \delta^2 f^2 \partial_t^{-1} \tilde{u} &=&0 \label{eq2}
    \end{eqnarray} \end{linenomath}
Again, the leading order gives $c_0^2= \alpha_1 \alpha_4 =\alpha_1 g (\rho-\rho_1).$ From the coupled system of equations one can eliminate $\tilde{u}$ by using an expansion like \eqref{JT} however taking into account all possible terms
\begin{equation}
  \tilde{u}=\frac{\alpha_4}{c_0} \eta + \delta^2 b_1\eta_{xx}+ \delta b_2 \frac{\eta^2}{2}+  \delta^2 b_3 \eta^3 +
  \delta^2 b_4 \partial_x  ^{-2}    \eta    \label{JT1}
\end{equation}
for some yet unknown coefficients $b_1,\ldots,b_4.$ The substitution of \eqref{JT1} in \eqref{eq1} leads to an equation for $\eta$,
keeping only terms up to order $\delta^2$:\begin{linenomath}
\begin{multline} \label{eq1prime}
\eta_t + c_0 \eta _x + \left[ \delta^2 \left(b_1\alpha_1+\frac{\alpha_2\alpha_4}{c_0} \right)\eta_{xx}   + \delta \left(\frac{\alpha_1 b_2 }{2}+\frac{\alpha_3 \alpha_4}{c_0}\right )\eta^2 \right.\\ \left.
+\delta^2 \left(\alpha_1 b_3 +\frac{\alpha_3 b_2}{2} -\frac{\beta_1 \alpha_4}{c_0}  \right) \eta^3 \right]_x
 +\delta^2 \alpha_1 b_4 \partial _x^{-1} \eta =0 \end{multline}\end{linenomath}

 The next step is in a similar fashion to eliminate $\tilde{u}$ from \eqref{eq2}. We need to substitute \eqref{JT1} in \eqref{eq2} however
there will appear $t-$ derivatives in various terms. In the non-leading order terms $\eta_t$ could be eliminated by using the expression
\eqref{eq1prime}. We also use the fact that in the leading order $\partial _t^{-1}=c_0^{-1} \partial_x ^{-1}.$ The result is
\begin{linenomath}
\begin{multline} \label{eq2prime}
\eta_t + c_0 \eta _x - \delta^2 \frac{c_0^2 b_1}{\alpha_4} \eta_{xxx}   + \delta \left(\frac{\alpha_3 \alpha_4}{c_0}-\frac{b_2 c_0^2}{\alpha_4}\right )\eta \eta_x\\
+\delta^2 \left(\frac{\alpha_3 b_2}{2} -\frac{\beta_1 \alpha_4}{c_0}-\frac{c_0 b_2}{3\alpha_4}\left( b_2 \alpha_1 + \frac{2\alpha_3 \alpha_4}{c_0}\right) \right) (\eta^3)_x -\delta^2\left( \frac{f^2}{c_0} + \frac{c_0^2 b_4}{\alpha_4} \right) \partial _x^{-1} \eta =0 \end{multline}\end{linenomath}

The two equations \eqref{eq1prime} and \eqref{eq2prime} of course need to coinside identically, and therefore there coefficients are equal. This leads to the following identities  (recall $c_0^2=\alpha_1 \alpha_4$)
\begin{equation}
    \begin{split}
        -\frac{c_0^2 b_1}{\alpha_4}=b_1\alpha_1+\frac{\alpha_2 \alpha_4}{c_0}\quad & \text{giving}\quad b_1=-\frac{c_0 \alpha_2}{2\alpha_1^2},\\
        \frac{\alpha_3 \alpha_4}{c_0}-\frac{b_2 c_0^2}{\alpha_4}= b_2 \alpha_1 + \frac{2\alpha_3 \alpha_4}{c_0}
        \quad & \text{giving}\quad b_2=-\frac{c_0 \alpha_3}{2\alpha_1 ^2},\\
               \alpha_1 b_3 +\frac{\alpha_3 b_2}{2} -\frac{\beta_1 \alpha_4}{c_0} \phantom{XXXXXXXXXXXX}& \\
               =\frac{\alpha_3 b_2}{2} -\frac{\beta_1 \alpha_4}{c_0}-\frac{c_0 b_2}{3\alpha_4}\left( b_2\alpha_1 + \frac{2\alpha_3 \alpha_4}{c_0}\right)  \quad & \text{giving}\quad  b_3= \frac{c_0 \alpha_3^2}{8\alpha_1^3}  ,\\
                \alpha_1 b_4=-\frac{f^2}{c_0} - \frac{c_0^2 b_4}{\alpha_4} \quad & \text{giving}\quad
        b_4=-\frac{f^2}{2c_0 \alpha_1}.
    \end{split}
\end{equation}
These coefficients completely determine the relation \eqref{JT1} giving $\tilde{u}$ in terms of $\eta.$ Moreover, the substitution of the coefficients in any of the equations  \eqref{eq1prime} or \eqref{eq2prime} leads to the following equation for $\eta:$
\begin{linenomath}
\begin{multline} \label{eq_final}
\eta_t + c_0 \eta _x +  \delta^2 \frac{c_0\alpha_2}{2\alpha_1} \eta_{xxx}
+ \delta \frac{3 c_0 \alpha_3 }{2\alpha_1} \eta \eta_x
-\delta^2 \frac{c_0}{8\alpha_1^2} \left(\alpha_3^2 + 8 \alpha_1 \beta_1  \right) (\eta^3 ) _x \\
-\delta^2 \frac{f^2}{2c_0} \partial _x^{-1} \eta =0 . \end{multline} \end{linenomath}
The coefficient in front of $(\eta^3 ) _x $ in terms of the physical parameters is
\begin{equation} -\frac{c_0}{8\alpha_1^2} \left(\alpha_3^2 + 8 \alpha_1 \beta_1  \right)= -\frac{c_0(\rho^2 h_1^4 +8\rho \rho_1 hh_1(h^2+h_1^2) + 14 \rho \rho_1 h^2 h_1^2 + \rho_1 h^4)}{8h^2 h_1^2(\rho_1 h + \rho h_1)^2}.\end{equation}

We observe that indeed, if $\alpha_3$ is of order of $\delta$ or smaller, then all non-leading order terms are of order $\delta^2.$
In terms of the $(X,T)$ variables introduced in the previous section, the leading order term $\eta_t + c_0 \eta _x =\delta^2 \eta_T$
is of the same order, $\delta^2.$

The equation with $\eta^3$ term (of type KdV-5 when $f=0$) appears already in \cite{DR1,Koop,KST} and more recently in the case with surface tension in \cite{GKKS}.

Finally, the very special case without quadratic nonlinearities is realised when $\alpha_3=0.$ Then the equation \eqref{eq_final} when $f=0$ is of mKdV-type for the real variable $\eta.$ Since the two densities are very similar, it is the difference $\Delta \rho=\rho-\rho_1$ that matters for the evaluation of
$\alpha_1.$ Elsewhere we can put $\rho=\rho_1,$ and hence $\alpha_3=0$ means $h=h_1.$ Then \eqref{eq_final} becomes
\begin{equation} \label{eq_special}
\eta_t + c_0 \eta _x +  \delta^2 \frac{c_0 h^2 }{6} \eta_{xxx}
-\delta^2 \frac{c_0}{h^2}  (\eta^3 ) _x -\delta^2 \frac{f^2}{2c_0} \partial _x^{-1} \eta =0  \end{equation}
with \begin{equation}c_0^2=\frac{gh\Delta \rho}{2\rho},\nonumber \end{equation} or
\begin{equation} \label{eq_special2}
\eta_T +\frac{c_0 h^2 }{6} \eta_{XXX}  - \frac{c_0}{h^2}  (\eta^3 ) _X - \frac{f^2}{2c_0} \partial _X^{-1} \eta =0 . \end{equation}

The mKdV equation (when $f=0$) is an integrable system, while the equation with a nonzero $f$ is not.

\section{Intermediate Long Wave approximation  }

In this section we study the equations of motion under the additional approximation that the wavelengths $\lambda$ are much bigger than $h_1$, i.e. the shallowness parameter will be taken as
\begin{equation}\delta={\frac{h_1}{\lambda}} \ll 1. \nonumber \end{equation} Noting that the wave number $k=2\pi/\lambda $ is an eigenvalue or a Fourier multiplier for the operator $D$ (when acting on monochromatic waves of the form $e^{ikx}$)  we make the following further assumptions about the scales:

1.  $\eta$ and $\tilde{u}$ are both of order $\delta;$

2. $h k=\mathcal{O}(1)$ and $h_1k=\mathcal{O}(\delta)$ i.e. $ h_1/h \sim \delta \ll 1$. This corresponds to a deep lower layer;

3. The physical constants $h_1,$ $\rho,$ $\rho_1,$ are $\mathcal{O}(1) ,$ and $f\sim \delta ^{3/2}$ (so that the $f^2$-term in the equation is of order  $\delta$).

Since the operator $D$ has an eigenvalue $k$, thus we shall also keep in mind that $hD=\mathcal{O}(1) $ and $h_1D=\mathcal{O}(\delta) .$
This regime is studied in more details in \cite{JCI} when of course $f=0.$ The Hamiltonian, expanded over $\delta$ is derived in \cite{JCI} and could be obtained again with an asymptotic expansion of the general Hamiltonian from \cite{CI2017}.
The Hamiltonian with terms up to order $\delta^3$ is \cite{JCI}
\begin{linenomath}
\begin{multline}
\label{H_ilw}
H_0[\eta,\mathfrak{u}]=\delta^2 \frac{h_1}{2\rho_1}\int_{\mathbb{R}} \tilde{u}^2 \,dx+\delta^2 \frac{\alpha_4}{2}\int_{\mathbb{R}}\eta^2 \, dx    -\delta^3 \frac{1}{2\rho_1} \int_{\mathbb{R}}  \eta \tilde{u} ^2   \,dx \\ - \delta^3 \frac{h_1^2\rho}{2\rho_1^2} \int_{\mathbb{R}} \tilde{u} \mathcal{T}_h \tilde{u}_x   \,dx
\end{multline} \end{linenomath}
where again the constant $\alpha_4=g(\rho-\rho_1).$
We notice that $H_0$ is of order $\delta^2$. The equations follow from the nearly-Hamiltonian formulation in the case of Coriolis force \eqref{H_Coriolis}, with the only difference that the Coriolis term with $f^2$ is now of order $\delta$:
\begin{eqnarray}
\label{ILW_1}
\eta_t + \frac{h_1}{\rho_1}\tilde{u}_x-\delta\frac{1}{\rho_1}(\eta\tilde{u})_x
-\delta \frac{\rho h_1^2}{\rho_1^2} \mathcal{T}_h  \tilde{u}_{xx}&=&0 \\
\label{ILW_2}
 \tilde{u}_t  + \alpha_4\eta_x -\delta\frac{1}{\rho_1}\tilde{u}\tilde{u}_x
+\delta f^2 \partial_t ^{-1} \tilde{u} &=&0.
\end{eqnarray}

The leading order terms (i.e. neglecting the terms with $\delta$ above) produce a system of linear equations with constant coefficients from where the speed(s) of the travelling waves (in the leading order) is
 \begin{equation} \label{c4BO}
c_0^2=\frac{h_1}{\rho_1}g(\rho-\rho_1) \end{equation}
and clearly corresponds to the limit of large $h $  in \eqref{c_0}.

With the Johnson transformation
\begin{equation}
  \tilde{u}=\frac{c_0 \rho_1}{h_1} \eta + \delta b_1  \mathcal{T}_h \eta_{x}+ \delta b_2 \frac{\eta^2}{2}+  \delta b_3 \partial_x  ^{-2}    \eta    \label{JT2}
\end{equation}
as before, we exclude the $\tilde{u}$ variable and obtain the two equations for $\eta:$
\begin{linenomath}
\begin{multline} \nonumber
\eta_t + c_0 \eta _x +  \delta \left(\frac{b_1 h_1}{\rho_1}-\frac{c_0\rho h_1}{\rho_1} \right) \mathcal{T}_h \eta_{xx}   + \delta \left(\frac{h_1 b_2 }{\rho_1}-\frac{2c_0}{h_1}\right )\eta \eta_x  +\delta  \frac{b_3 h_1}{\rho_1}\partial _x^{-1} \eta =0,
\\
\eta_t + c_0 \eta _x -  \delta \frac{b_1 h_1}{\rho_1} \mathcal{T}_h \eta_{xx}   - \delta \left(\frac{h_1 b_2 }{\rho_1}+\frac{c_0}{h_1}\right )\eta \eta_x  -\delta \left( \frac{b_3 h_1}{\rho_1} + \frac{f^2}{c_0}\right) \partial _x^{-1} \eta =0.
\end{multline}
\end{linenomath}

From the comparison of the corresponding coefficients we determine the quantities
\begin{equation}
    b_1=\frac{c_0 \rho}{2}, \quad b_2=\frac{c_0 \rho_1}{2h_1^2}, \quad b_3=- \frac{f^2\rho_1}{2 c_0 h_1}
\end{equation}

which leads to the following equation for $\eta:$
\begin{equation}
    \eta_t + c_0 \eta _x -  \delta\frac{c_0\rho h_1}{2\rho_1}  \mathcal{T}_h \eta_{xx} - \delta \frac{3c_0}{2h_1}\eta \eta_x  -\delta  \frac{f^2}{2c_0}\partial _x^{-1} \eta =0.
\end{equation}

With an appropriate Galilean transformation $X\rightarrow (x-c_0t),$  $T\rightarrow \delta t$  the equation transforms into
\begin{equation} \label{ilwf}
    \eta_T - \frac{c_0\rho h_1}{2\rho_1}  \mathcal{T}_h \eta_{XX} -  \frac{3c_0}{2h_1}\eta \eta_X  - \frac{f^2}{2c_0}\partial _X^{-1} \eta =0.
\end{equation}

We note that the speed $c_0$ depends only on the depth of the upper layer $h_1.$ The $h$-dependence comes only from the term $\mathcal{T}_h.$ For $c_0$ there are two possible choices, corresponding to the left or the right-running waves. For the two choices the equations \eqref{ilwf} are of course different. The obtained equation \eqref{ilwf} extends the integrable Intermediate Long Wave (ILW) equation \cite{Jo,Ku,Chen}, there are more details for example in \cite{BonnaLannesSaut,JCI}.

In the limit $h\to \infty$ the operator $\mathcal{T}_h=-i \coth (hD)$ becomes the
Hilbert transform operator $\mathcal{H}=-i\,\text{sign}(D)$ or $\mathcal{T}_h \partial_x \rightarrow |D|=|\partial_X| .$ Then \eqref{ilwf} becomes an extension of the Benjamin-Ono (BO) equation:

\begin{equation} \label{bof}
    \eta_T - \frac{c_0\rho h_1}{2\rho_1}  |\partial_X| \eta_X -  \frac{3c_0}{2h_1}\eta \eta_X  - \frac{f^2}{2c_0}\partial _X^{-1} \eta =0.
\end{equation}

The BO equation \cite{Benj,Ono} like KdV and ILWE is a classical example of an integrable equation solvable by the inverse scattering method \cite{KM98}. For more information about these models one can refer to \cite{FA,BonnaLannesSaut,CI2019} and the references therein.

The models with Coriolis term \eqref{bof}, \eqref{ilwf} and the others are not integrable, however the solitary wave solutions could be treated in the framework of the theory of soliton perturbations, as well as with other methods.

\section{Including the $y$-dependence} \label{y-depe}

So far we have ignored the $y$-dependence of the physical quantities, and there are physical situations where this is justified. However, from \eqref{y-dep} it is clear that the $y$ derivative, or, rather the operator $-i\partial_y$ and its eigenvalue $k_y$ are of the same order as $f$, that is $\delta^{n/2}$  where $n=4$ for the KdV models and $n=3$ for the ILWE and BO models. Now we reintroduce this dependence through the following Lemma:
\begin{lemma}\label{lem} The approximate model equations for small amplitude and long or intermediate long wave regimes have the following general form
\begin{equation}\label{lint}
    \eta_t +c_0\eta_x +\delta^{n-2}\left[\frac{R(D)}{2c_0}\eta + \left(\frac{c_0 }{2}\partial_y^2- \frac{f^2}{2c_0 }\right) \partial_x^{-1} \eta \right] + \text{nonlin. terms}=0,
\end{equation} where $R(k)$ is an appropriate odd function.
\end{lemma}
Proof: Since $\varphi_y \ll \varphi_x $ from the symmetry of the original equations it is clear that when extending the gradient there will be no $\partial_y$ contributions in the nonlinear part of the equation, since $\partial_y$ would not contribute to the leading order of the nonlinear part. The contribution of $\partial_y$ in the dispersive part of the equation could be determined from the extension of the dispersion law. The dispersion law in principle is given by some model-related odd function $R(k)$ - like in \eqref{dl} where $R(k)=-\frac{c_0^2 \alpha_2}{\alpha_1}k^3,$ and in our situation, in addition - a Coriolis contribution, given by the $f^2$ term,
\begin{equation}
    \tilde{\omega}^2(k)=c_0^2k^2 +\delta^{n-2}\left[k R(k) + f^2 \right]
\end{equation}
The leading order term could be extended in the two-dimensional case noting that  $k^2 \rightarrow \Vec{k}^2= k^2+k_y^2$, or taking into account the scaling
\begin{equation}\delta^2k^2 \rightarrow \delta^2  k^2+\delta^n k_y^2= \delta^2(k^2 + \delta^{n-2}k_y^2)\nonumber \end{equation} hence
\begin{equation}
    \tilde{\omega}^2(k,k_y)=c_0^2k^2 +\delta^{n-2}\left[c_0^2 k_y^2 + k R(k) + f^2 \right]
\end{equation}
The evaluation of the square root up to $\delta^{n-2}$ gives
\begin{equation}
    \tilde{\omega}(k,k_y)=c_0k +\delta^{n-2}\left[\frac{R(k)}{2c_0 }+ \frac{c_0 k_y^2}{2k}+ \frac{f^2}{2c_0 k} \right].
\end{equation}
The equation for $\eta$ is of the form $\eta_t=i \tilde{\omega}(D,-i\partial_y)\eta$+ nonlinear terms, that is \eqref{lint}. $\qed$

Therefore the addition to the model equations in each case is \begin{equation}\delta^{n-2} \left(\frac{c_0 }{2}\partial_y^2- \frac{f^2}{2c_0 }\right) \partial_x^{-1} \eta . \nonumber \end{equation}  We note that the $y$-derivatives term and the Coriolis term are both of the same order.

 The main results, following from the Main Theorem and \eqref{H1}-\eqref{H2} with the $y$-dependence extension from Lemma \ref{lem} therefore could be summarised into the following corollaries about the approximate models describing the time-evolution of the internal wave in two dimensions:
\begin{corollary}
 (a) In the case of small amplitude (of order $\delta^2$) and long-wave regime, the approximation gives the KP - Ostrovsky type equation
\begin{equation}\label{KdV_final_y}
  \eta_t + c_0 \eta_x + \delta^2 \frac{\alpha_2 c_0}{2 \alpha_1} \eta_{xxx}+ \delta^2 \frac{3c_0\alpha_3}{4\alpha_1} (\eta^2)_x
  +\delta^2 \left(\frac{c_0 }{2}\partial_y^2- \frac{f^2}{2c_0 }\right) \partial _x^{-1} \eta =0;
\end{equation}
(b)  In the case of small amplitude (of order $\delta$) with parameter $\alpha_3 $ of order $\delta$ or smaller, and long-wave regime, the approximation gives the mKdV - KP - Ostrovsky type equation
\begin{linenomath}
\begin{multline} \label{KP_final_y}
\eta_t + c_0 \eta _x +  \delta^2 \frac{c_0\alpha_2}{2\alpha_1} \eta_{xxx}
+ \delta \frac{3 c_0 \alpha_3 }{2\alpha_1} \eta \eta_x
-\delta^2 \frac{c_0}{8\alpha_1^2} \left(\alpha_3^2 + 8 \alpha_1 \beta_1  \right) (\eta^3 ) _x \\
+\delta^2 \left(\frac{c_0 }{2}\partial_y^2- \frac{f^2}{2c_0 }\right) \partial _x^{-1} \eta =0 . \end{multline}\end{linenomath}
\end{corollary}
 By taking formally $f=0$ in \eqref{KdV_final_y} one can recover the corresponding Kadomtsev-Petviashvili (KP)-type equations, \cite{KP,ZMNP,Johnson_Book}.
\begin{corollary}
 (a) In the case of small amplitude (of order $\delta$)  and intermediate long wave regime, the approximation gives the ILW - Ostrovsky type equation
\begin{equation}
    \eta_t + c_0 \eta _x -  \delta\frac{c_0\rho h_1}{2\rho_1}  \mathcal{T}_h \eta_{xx} - \delta \frac{3c_0}{2h_1}\eta \eta_x  +\delta \left(\frac{c_0 }{2}\partial_y^2- \frac{f^2}{2c_0 }\right) \partial _x^{-1} \eta =0,
\end{equation}
and,
\\
 (b) In the limit $h\to \infty$, the previous model becomes the BO-Ostrovsky type equation:
\begin{equation}\label{BO_y_final}
    \eta_t + c_0 \eta _x -  \delta\frac{c_0\rho h_1}{2\rho_1}  |\partial_x| \eta_{x} - \delta \frac{3c_0}{2h_1}\eta \eta_x  +\delta \left(\frac{c_0 }{2}\partial_y^2- \frac{f^2}{2c_0 }\right) \partial _x^{-1} \eta =0.
\end{equation}
\end{corollary}

\section{Conclusions}

We have presented a consistent approach for the derivation of the simplified long-wave and intermediate long-wave models \eqref{KdV_final_y}-\eqref{BO_y_final} based on the extension of the Hamiltonian approach in the irrotational case \cite{Craig2,Craig}. A similar Hamiltonian approach has been successful for treating internal waves with shear currents \cite{CI2017,CI2019,CI,CI3,CIM,I17} and hopefully could work in situations where other effects affect the long - wave or the intermediate long-wave propagation.  For surface waves the derivation is analogous, only the Hamiltonian $H_0$ is the corresponding Hamiltonian for the propagation of surface waves.

The obtained approximate models are integrable only in the special case of $f=0,$ and they are not integrable in the situation that includes the Coriolis forces. Nevertheless the structure of the equations usually allows several conservation laws (like ``mass'' and ``energy'' ) and also allows the application of the soliton perturbation theory to the soliton solutions of the corresponding integrable equations \cite{GI}. In addition, the structure of perturbed integrable equations has certain advantages in the development of numerical methods for these equations.

\section*{Acknowledgements}

The authors are thankful to two anonymous referees for their valuable suggestions.

%For acknowledgements section, please don't number the section, please begin it with \section*{Acknowledgements}
%\section*{Acknowledgments} R.I. is partially supported by the Bulgarian National Science Fund, grant K$\Pi$ -06H42/2
%from 27.11.2020.

% You may incorporate your references as follows in your main tex file.
% Using BibTex is not recommended but can be handled.
%

\medskip
% The data information below will be filled by AIMS editorial staff
Received  October 2021; revised  December 2021.
\medskip

\end{document}